# Natural Language Embeddings of Synthesis and Testing conditions Enhance Glass Dissolution Prediction

Sajid Mannan[1], K. Sidharth Nambudiripad[1], Indrajeet Mandal[2], Nitya Nand Gosvami[3], N. M. Anoop Krishnan[1,4,*]

[1]*Department of Civil and Environmental Engineering, Indian Institute of Technology Delhi, Hauz Khas, New Delhi-110016, India*
[2]*School of Interdisciplinary Research, Indian Institute of Technology Delhi, Hauz Khas, New Delhi 110016, India*
[3]*Department of Materials Science and Engineering, Indian Institute of Technology Delhi, Hauz Khas, New Delhi-110016, India*
[4]*Yardi School of Artificial Intelligence, Indian Institute of Technology Delhi, Hauz Khas, New Delhi 110016, India*

[*]*Corresponding author: NMAK (krishnan@iitd.ac.in)*

**ABSTRACT**

Long-term chemical durability of glass, crucial for immobilizing nuclear waste, is governed by glass properties such as composition, surface geometry, as well as external factors like thermodynamic conditions and surrounding medium. Despite decades of research, there are no models that account for these intrinsic and extrinsic factors to predict the dissolution rates of glass compositions. To address this challenge, we evaluate the role of natural language embeddings capturing the synthesis and testing conditions in enhancing the predictability of glass dissolution. Evaluating the approach on hand-curated ~700 datapoints extracted from the literature, we reveal that the machine learning (ML) model including natural language embeddings (NLP-ML) outperforms classical ML model in predicting glass dissolution rate. Furthermore, we developed a generalizable ML model by transforming the compositional features to structural descriptors of glass alongside NLP-derived features, enabling extrapolation capability to glass compositions with completely new elements absent in the training data. Evaluating this model on a completely new dataset of glass compositions with 34 chemical components in contrast to the training dataset that had only 28 components, we demonstrate that the model indeed exhibits generalizability to glass compositions that are out-of-distribution. Altogether, this integrated approach offers a pathway towards high-fidelity glass dissolution prediction and accelerate the discovery of novel glass compositions with tailored durability for sustainable nuclear waste management.

# 1. INTRODUCTION

Effective immobilization of radioactive waste is a critical global concern. Glass has emerged as a highly effective material for immobilizing nuclear waste[1] due to its high chemical and physical stability and ease of production. Moreover, the glass matrix can readily incorporate many elements in nuclear waste, and such a glass needs to exhibit long-term stability and durability. Currently, high-level nuclear waste is vitrified with borosilicate glasses and allowed to cool in the atmosphere before being stored in geological repositories hundreds of meters below the surface to prevent human exposure. Nonetheless, this storage method introduces the possibility of underground water exposure, leading to glass corrosion[2] and water contamination. Given the thermodynamic instability[3–5] of glasses, their interaction with water can prompt transformations into more stable chemical compounds. Various intrinsic and extrinsic factors, including temperature, pH of the medium, and glass composition, influence glass corrosion. It occurs under non-equilibrium conditions through complex and interdependent pathways[4,6–16] such as network hydrolysis, ion exchange, and the formation of secondary phases. Thus, comprehending the degradation mechanisms of glasses presents a significant challenge. Therefore, a thorough understanding of glass dissolution is crucial for the long-term storage of nuclear waste glasses.

The Glass Reactivity with Allowance for the Alteration Layer (GRAAL) model[17–19] currently serves as the standard mathematical model for modeling glass dissolution behavior and designing protocols to manage the disposal of nuclear waste[20,21]. However, the lack of understanding regarding long-term glass dissolution behavior in various kinds of aqueous environmental conditions makes it challenging for a purely mechanical model to offer long-term solutions for the wide variety of glasses in the compositional space[22]. Recently, AI/ML have aided efforts to develop data-driven approaches as potential alternatives to other empirical and mechanism-based models[20,22]. In the context of glasses, researchers have made remarkable attempts to predict various properties using machine learning[23–26]. There have also been attempts to interpret the underlying workings of these ML models, which are otherwise black boxes in nature[26–29].

In context of glass dissolution kinetics, Krishnan et al.[22] exemplified the usefulness of ML models with their work on predicting the dissolution rate of silicon for ternary sodium aluminosilicate glasses. Results showed the capability of the predictive ML models to capture the underlying relationship between the dissolution rate and the glass composition, which agreed with the experimental values. Furthermore, Lillington et al.[20] expanded the work by conducting research on the efficacy of ML models in predicting the dissolution rate of radioactive waste glasses. They considered complex, multi-component glasses and trained ML models by compiling different datasets for static and dynamic leach tests. However, there are two significant issues with these kinds of models. (i) These ML models are still *blind* and require vast amount of data to improve their predictive performance[30,31]. (ii) These ML models have the limitation of generalizability and are ineffective in explaining the relationship between the dissolution kinetics and the broad spectrum of real-world conditions that these glasses must endure[32]. They are restricted to the actual set of components used for training, which results in a lack of interpretability.

A major reason for these could be attributed to the fact that the properties of materials, and glasses in particular, are extremely sensitive to the synthesis and testing conditions. While most of the models rely on composition or other such features as inputs, they completely ignore the testing conditions during to their unstructured and variable nature; for instance, synthesis conditions cannot be represented as a tabular data. To this end, Sasidhar et al.[30] demonstrated that combining textual information as word embeddings can enable improved design of corrosion-resistant alloys. Specifically, they employed Recurrent Neural Networks (RNNs) to leverage textual data, which were utilized in their deep learning framework to predict the corrosion-resistant alloy design. While results demonstrated that fusion of textual data with numerical data can enhance the predictive accuracy, limited efforts are made to exploit the advances in language model development, especially the transformer-based models, that enable a more nuanced and semantically rich representation of the synthesis and testing conditions.

Here, we propose the use of transformer-based domain specific language model, that exploits the synthesis and testing conditions of materials, represented as textual data to predict the dissolution of glasses. Specifically, the experimental information relevant to glass dissolution experiments are transformed into numerical form using MatSciBERT[33] and combined with the original numerical data (such as Composition, pH, and temperature), to evaluate their effect on the prediction of glass dissolution rate. Results using this model, named as NLP-ML model, demonstrate that, coupling of textual features with numerical features improves the prediction accuracy. Furthermore, to enhance the generalizability of NLP-ML model, we convert the compositions into various fundamental physical and chemical descriptors. The composition-converted-descriptors along with NLP generated features are used to train the descriptor-based NLP-ML model. Results demonstrate that the descriptor-based NLP-ML model exhibits reasonable accuracy in comparison to the composition-based NLP-ML model, highlighting the effectiveness of the chosen descriptors in capturing the compositional information. Furthermore, we evaluate the descriptor-based NLP-ML model on unseen glass compositions and validate the generalization capabilities of the model. Finally, we employ a model-agnostic, game-theoretic approach called SHapley Additive exPlanations (SHAP)[34,35] to interpret the role of the individual input features in the decision-making processes of the descriptor-based NLP-ML model and unravel the underlying input-output relationship.

## 2. METHODS
### 2.1. Dataset preparation
The dissolution dataset used to train the NLP-ML models was manually curated, consisting of 693 data points specifying the leaching rates of glasses with respect to silicon (Si) release from numerous published works[36–52]. Special attention was paid while preparing this dataset to ensure the incorporation of a wide compositional space, different pH values, and multiple temperature conditions. Numerical data corresponding to composition, pH and temperature were collected from tables and plots (plot data was collected using PlotDigitizer[53]). To ensure data consistency, all glass compositions were normalized, and it was made sure that the final glass composition sum added up to 100%. Note that all features were retained in numerical form and glass compositions were taken using mole percentage. Duplicate entries were systematically eliminated from the dataset. When multiple dissolution rate values were found for a given input (Composition, pH, and Temperature), outliers were identified as those values lying beyond ±2.5% of the mean of all dissolution rates for that input. These outliers were subsequently removed, and the mean value of the remaining dissolution rate values was considered the dissolution rate of that input. This rigorous protocol resulted in a dissolution rate dataset with ~530 data points comprising 53 unique glass components (taken in mole percentage) along with a broad temperature range (4°C to 300°C) as well as pH in both acidic and basic regions (1 to 13). Note: All the glasses, whose rate values were measured above 100°C temperature, were tested in controlled environments with highly sophisticated methods as these measurements were carried out in high-pressure conditions. The experimental procedure either involved using a mixed-flow hydrothermal reactor setup[39,51,54] or the accelerated leach test of glasses[47] under high pressures (~17 bar). This dataset consisted of data with respect to both static and dynamic leach tests.

## 2.2. Addition of Natural Language Input Features

Textual data contain rich, yet unstructured representation, of the synthesis and testing conditions while preparing the sample and measuring the property. MatSciBERT[33], a materials science domain-specific language model, was used for incorporating information regarding the experimental procedure and processing conditions. Further, the textual data, in the form of vector embeddings, was combined with the numerical features[33,55]. In this approach, textual data from the *Experimental Methods* section of the selected research papers (which made up the dataset) were separately extracted and retained. The curated textual database comprised 21 entries. The collected data was tokenized using MatSciBERT, which uses a WordPiece tokenizer[33,56,57]. Note that the original feature from MatSciBERT is a 768-dimensional embedding which is substantially larger than the context window of the model.

In tokenization, the text is segmented into sub-word units called tokens. Note that the upper limit on the number of tokens for MatSciBERT in a sequence is 512 tokens. Hence, the collected textual data was rephrased and summarized such that after tokenization, the number of tokens did not exceed 512. It was ensured that information highlighting glass preparation methods, thermal-mechanical heat treatments, experimental equipment used in the dissolution experiments, intrinsic glass properties (wherever mentioned), surface area calculation methods, testing conditions (such as the leachants used, pressure magnitudes during testing, static/dynamic leach tests etc.) were strictly retained in the final textual input to the tokenizer.

To ensure the balance between numerical feature and textual feature and avoid large number of textual feature, we employed Uniform Manifold Approximation and Projection (UMAP)[58,59] to reduce the dimensionality of the textual embeddings. In this approach, the local and global structure of the 768-dimensional embedding space is retained in a lower, $x$-dimensional space ($x < 768$), with the help of fuzzy simplicial complexes. These are high-dimensional representations of the topological structure of the embedding space. A high dimensional fuzzy simplicial complex is constructed for each $n \times 768$ dimension embedding. The algorithm finds the *k* nearest neighbors for each of the *n* datapoints and connects each of them with its neighbor. Each such connection is assigned a value between 0 and 1 denoting the strength of the connection. This value is dependent on the density of the local neighborhood of the datapoint being considered. Following this, the lower, *x*-dimensional space is created and the *n* datapoints are arranged in the lower dimensional space as another fuzzy simplicial complex. The algorithm minimizes the differences between the high-dimensional and lower-dimensional simplicial complexes by optimizing a cross-entropy loss function with an optimization algorithm such as Stochastic Gradient Descent (SGD)[60,61].

## 2.3. Trustworthiness of the input

It is important to ensure that the reduced representation of the NLP features retain relevant information of the synthesis and testing conditions. To this end, we use the Scikit-learn[62] Trustworthiness score to calculate the appropriate number of components to retain important NLP features. Trustworthiness score highlights the effectiveness of preserving the local neighborhood relationships of the datapoints from the ground-truth high-dimensional embedding space. The mathematical formula for calculating the Trustworthiness is as follows:

$$T(k) = 1 - \frac{2}{nk(2n-3k-1)} \sum_{u=1}^{n} \sum_{w \in N_k(u)} max(0, r(u, \boldsymbol{w}) - k)$$

*n* = number of datapoints (or tokens)
*k* = number of neighbors considered
*r(u, w)* = index of point *w* in sorted list of distances from point *u* in the lower dimensional space
$N_k(u)$ = Set of k nearest neighbors in the high dimensional space.

To consider the trade-off between the preservation of the local and global structure in the lower dimensional space, and to ensure optimal model training, trustworthiness scores were calculated for all 21 *n*×768 dimensional embeddings, in an iterative manner, starting from a low value of 10, for the number of retained components. After each iteration, this value was increased. The favorable number of components to retain, was selected based on the following heuristic: *Trustworthiness score is tracked across every iteration corresponding to a particular component value. The iterative process stops after the trustworthiness score reaches a value ≥ 0.7. In cases where the trustworthiness score fails to reach the threshold of 0.7, a patience level of 5 iterations is set, and the favorable number of components is selected as the component value corresponding to the highest trustworthiness score.*

## 2.4. Physics and Chemistry-based Descriptors

The extrapolation capability of the descriptor-based NLP-ML model is owed to the transformation of the compositional feature space into quantitative descriptors. These descriptors capture the physical and chemical attributes of the glass structure and aid in providing comprehensive, quantitative information regarding chemical components that make up these glasses, which are beyond the distribution of the training data. We use the same twelve descriptors for the conversion of oxide components as Bishnoi et al.[63] used in their work. Note that all these descriptors are handcrafted and based on domain knowledge considering the following pivotal aspects: (i) The respective percentages of oxygen, network formers, and network modifiers significantly affect the properties of oxide glasses. (ii) The cationic strength and, consequently, the glass property are highly dependent on the valence of the formers and modifiers. (iii) Atomic properties such as the atomic mass, the atomic radii, and the atomic number are responsible for the packing of the glass structure, which affects other properties. Finally, (iv) the local charge distribution is governed by the polarizability of the glass, which is also known to affect material properties.

## 2.5. Unseen Test Dataset and the Jaccard Distance Metric

The manually curated separate unseen test data consists of 32 silicon initial dissolution rate datapoints, collected at various pH and temperature values, with respect to two glasses — International Simple Glass (Fournier et al.[64]) and P0798, a Japanese simulated high-level waste glass (Inagaki et al.[65]) . Temperature values in this dataset range from 25 °C to 90 °C and pH values range from pH 3 to pH 11.3. Before assessing the performance of the descriptor-based NLP-ML model, the dataset was processed in a manner similar to the one used to process the training dissolution dataset. Compositions were collected in mole percentage and thereafter converted to the quantitative descriptors.

Jaccard Distance is a metric used to quantify the dissimilarity between two sets. Mathematically, Jaccard Distance between two sets A and B is calculated as follows:

$$J(A, B) = 1 - \frac{A \cap B}{A \cup B}$$

In the context of the dissolution datasets presented in this study, the Jaccard Distance was used to quantify the dissimilarity between the separate unseen dataset and the training data. The set of chemical components which make up the corresponding glasses' structures were taken as the input sets for calculating the Jaccard Distance. This metric broadly quantified the chemical complexity differences, based on presence/absence of chemical components, between the glasses in both datasets and aided in establishing the usefulness of the separate unseen test dataset as an evaluation benchmark.

## 2.6. ML Model Development

To train the ML models, first, we randomly shuffle the processed datasets and split the data in an 80:20 ratio. These become the training and test datasets, respectively. For the composition-based NLP-ML model, glass compositional compounds, pH, temperature and transformed textual features are independent variables (for the standard model, all parameters except transformed natural language features are given as input), and dissolution rate becomes the dependent variable. For the descriptor-based NLP-ML model, the compositional inputs are first converted into descriptors. Subsequently, the composition-converted-descriptors, pH, temperature and transformed textual features are given as input parameters, to predict the dissolution rate of glasses. The Standard Scaler library[66] is used to scale the feature values based on their respective mean and standard deviation before performing the model training. Furthermore, a 10-fold cross-validation[67] approach is enforced for the optimal fit of hyperparameters. These hyperparameters are optimized using the Bayesian optimization library - Optuna[68]. In the case of Artificial Neural Network, these hyperparameters are epochs, n_layers, activation, opt, drop, and drate. The final model is chosen based on the best $R^2$ score on the validation fold, as the test data are kept hidden to avoid data leakage[69]. The details of all the hyperparameters are given in supplementary information (Table S4). In this study, we utilize several Python packages. PyTorch[70], an open-source machine learning library, is employed to develop a feed-forward multi-layer perceptron (MLP) and a tree-based eXtreme Gradient Boost (XGBoost) model. For pre-processing and data visualization, we use NumPy[71], Pandas[72] and Matplotlib[73]. All the codes and datasets utilized in this study are available on GitHub https://github.com/M3RG-IITD/NLP-ML.git.

**2.7. SHAP for tree ensemble**
SHAP (SHapely Additive exPlanations)[34,35] is a unified game-theoretic framework employed to elucidate the impact of the input features on the ML model output. In this approach, the feature values are considered the players, while the model output is conceptualized as the game itself. The importance of each feature is assessed by quantifying the prediction errors through perturbations in the feature value, which gives us profound insights into the decision-making processes of the machine learning model. Mathematically, it is defined as:

$$\phi_i = \sum_{S \subseteq Fi} \frac{|S|!\,(|F| - |S| - 1)!\,|F|!}{|F|!} [f_{S \cup i}(x_{S \cup i}) - f_S(x_S)]$$

where F denotes the set of all features, S denotes a subset of features, $f_{S \cup i}$ denotes a model trained with the feature present, $f_{S\{i\}}$ denotes a model trained with the feature withheld, and $x_S$ denotes the input features in subset S. Note that a larger prediction error denotes greater importance of the particular feature, while a smaller prediction error indicates lesser importance.

SHAP provides model-agnostic approximation techniques like KernelSHAP and model-specific ones like DeepSHAP, MaxSHAP, and LinearSHAP. Depending upon the task, we can opt for any explainers to calculate each feature's SHAP values. Here, we used DeepSHAP which leverages the internal structure of neural network models to explain model predictions by approximating Shapley values. It makes use of DeepLIFT[74], an attribution backpropagation method to track how the output is affected by the changes in input feature values and their propagation through the network. It makes use of baseline predictions and quantifies the role of each input feature in changing the output from its expected value. This produces feature attributions with desirable Shapley properties like additivity and consistency. SHAP also offers different kinds of plots for visualizing the model interpretation, such as bar, violin, decision, text, waterfall, and bee-swarm plots. In this study, we visualized the SHAP values for the descriptor-based NLP-ML model using a bar plot and a bee-swarm plot. Bar plots display the average impact of each feature on the model output. Bee-swarm plots display the contribution of each feature towards the model output (in order of priority, that is, most important to least important) that played a crucial role in the decision-making processes of the model. Both plots display the contribution of those features as a function of the feature value.

## 3. RESULTS AND DISCUSSION

First, we discuss the transformation of the unstructured textual information into features. Figure 1(a) shows the conversion of textual data detailing the experimental methodology and processing conditions into a compact form, which is used as input to the tokenizer. MatSciBERT[33], a BERT-based language model pretrained on materials sciences corpus, was used as the tokenizer to generate $n \times 768$ dimensional embeddings ($n$ = number of tokens in a single textual input) so that the domain-specific terminologies are retained, and the generated embeddings are more semantically meaningful. Following this, the features were further reduced using UMAP and the relevant features were selected using Trustworthiness score (see Fig. 2 and Methods for details). We observe that the mode of the distribution of the favorable components to be retained for each textual input is 10 (see Fig. 1(b)). Accordingly, the 10 components of the natural language embeddings retained was concatenated with other features such as compositions, pH, and temperature to predict the dissolution rate.

Figure 2 shows the overall pipeline for training the classical ML and NLP-ML models. The textual data is transformed into numerical representations called vector embeddings using a language model. It is worth noting that descriptions of glass preparation methods, thermal-mechanical heat treatment as well as certain intrinsic glass properties (wherever mentioned) were extracted and retained in the final curated textual database. For example, sentences such as "*For dissolution experiments, glass was dried at 60°C, hand-ground, and sieved to 50–125 µm, then ultrasonically cleaned in deionized water and acetone.*" and "*Quantities of 100 g of glass were mixed in an ethanol suspension in a ball mill for 4 hours, then dried, pelletized, and melted in platinum crucibles at 1600°C using a Borel MO 1700 high-temperature furnace*" and "*The glass density was $2.59 \times 10^3$ kg m$^{-3}$.*" were retained in the final database.

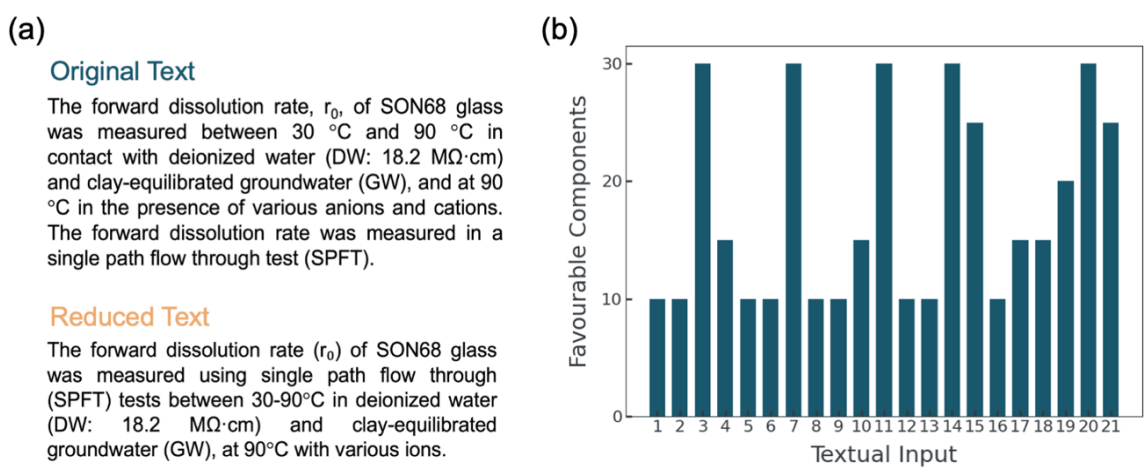

*Figure 1. (a) The conversion of original textual data to a reduced form for tokenization (b) Plot illustrating the selection of textual features to be added as natural language input features with the percentage of explained variance as a function of number of textual features.*

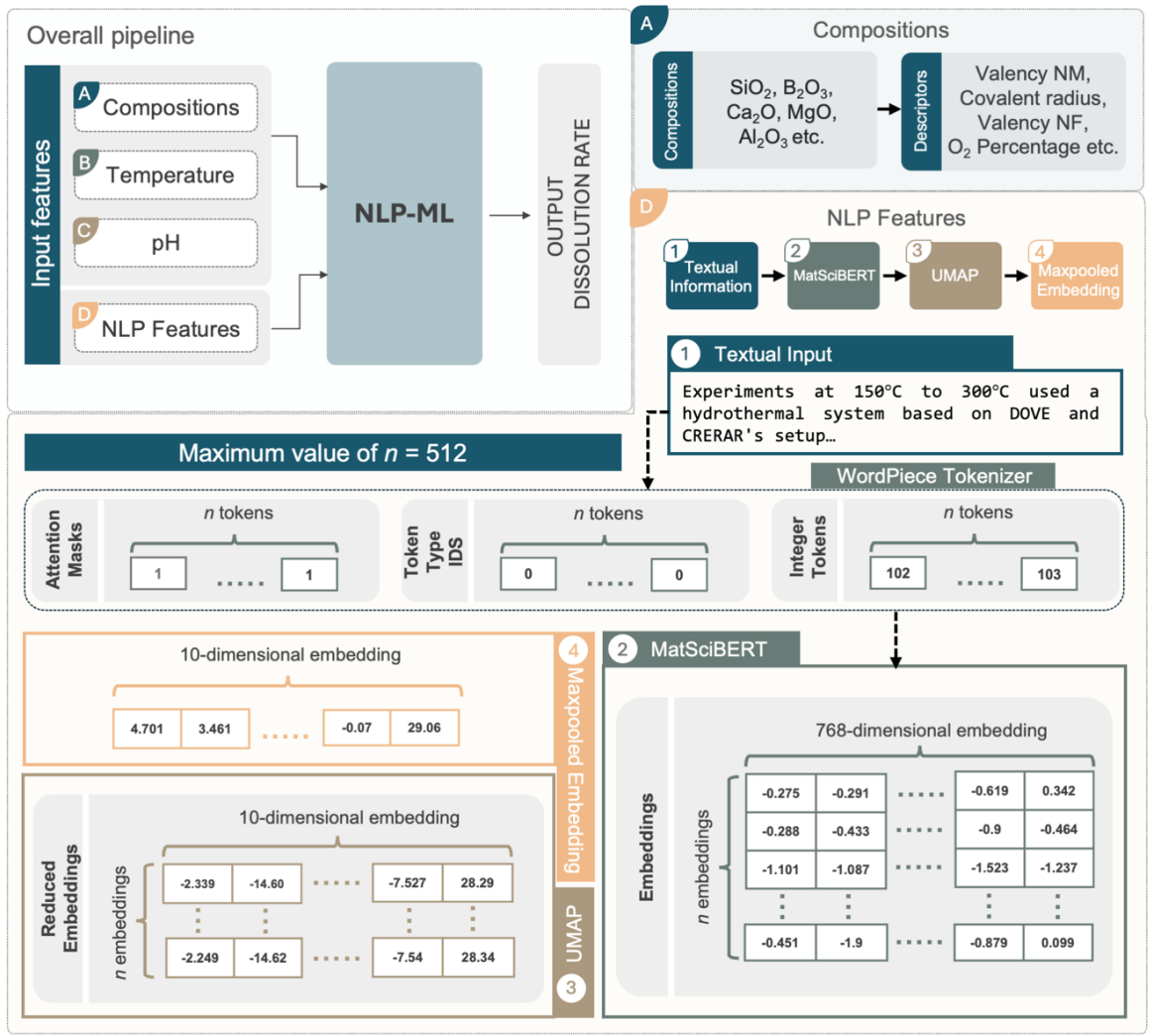

*Figure 2. Overview of NLP-ML Model for Predicting Dissolution Rates. The inset illustrates the NLP features creation from unstructured textual embedding using MatSciBERT.*

### 3.1. Dataset Visualization

Firstly, we analyze the dissolution dataset. The dataset consists of leaching rate data with respect to silicon release. Figure 3(a) shows a bar chart comparing the glass compositions associated with each feature in the dataset. Notably, the ML model comprises 53 compositional features. Figure 3(b) illustrates the distribution of the number of chemical elements in each glass composition across the dataset, indicating how many glasses consist of five, six, seven components, and so on. The plot shows that the curated dissolution data consists of a variety of glasses, from moderately complex to highly complex, containing between 5 and 28 components. Figure 3(c) presents a histogram of the dissolution rates in the dataset. The dissolution rate in the dataset, spans a wide numerical range, from approximately -12 to 11 *collected as $\log_e(rate)(g/m^2/day)$*, thus encompassing a broad spectrum of property values. Note: all plots also show the train-test split in the dissolution dataset.

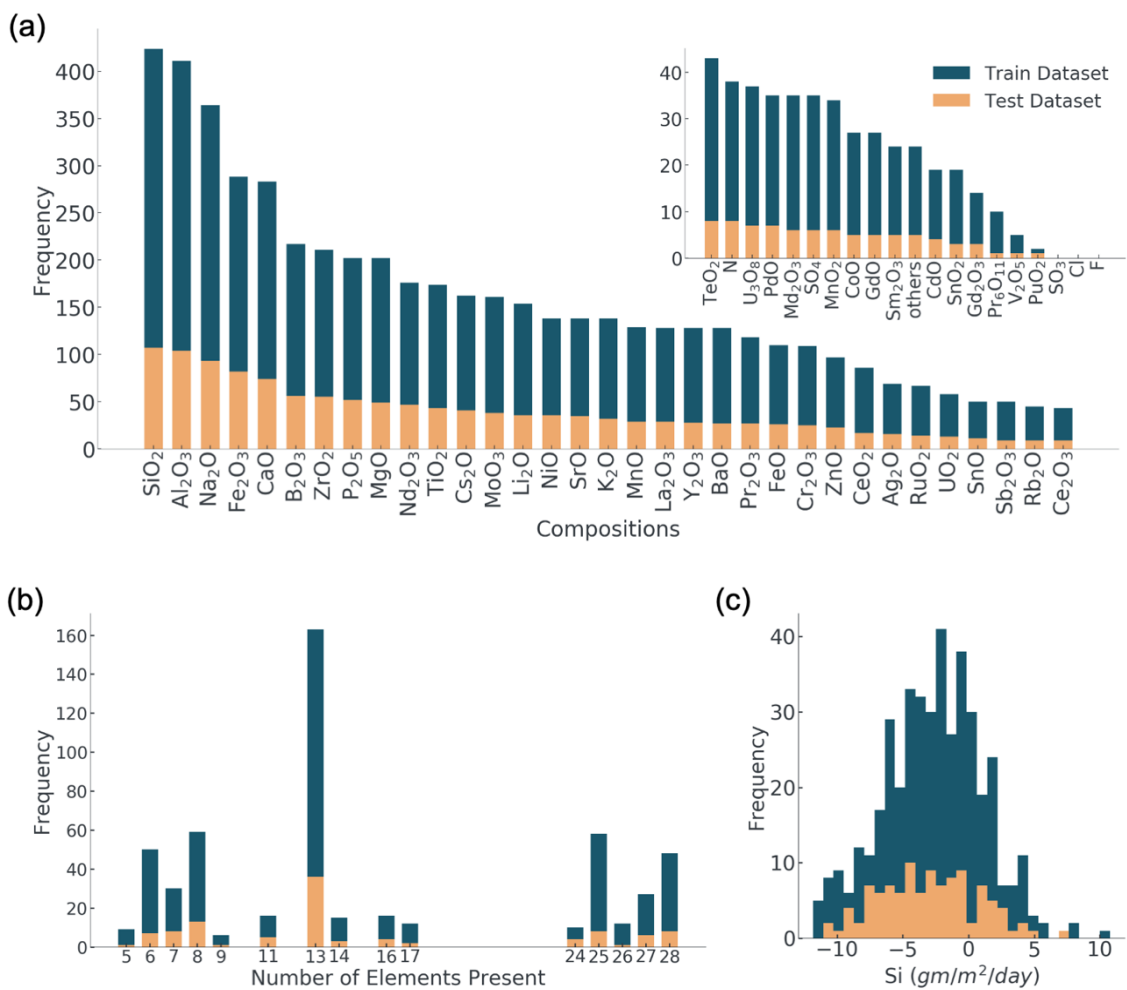

*Figure 3.* Visualizing the dissolution dataset: (a) Frequency of glass components in the dissolution dataset. (b) Complexity of glasses in the dissolution dataset. (c) Histogram of dissolution rate for the dissolution dataset.

### 3.2. Dissolution Rate Prediction

We trained two different ML models—tree-based eXtreme Gradient Boost (XGBoost)[75] and feed-forward multi-layer perceptron (MLP) to assess the capability of the models in capturing inherent composition-property relationships. Note, while XGBoost achieves a performance comparable to NN model, the addition of NLP features does not lead to a substantial improvement in its accuracy (see Supplementary Information Fig. 23-24). This behavior can be attributed to the inherent structure of tree-based models, wherein predictions are determined by recursive partitioning of the feature space into leaf nodes. In high-dimensional settings, such models often tend to assign relatively low importance to correlated features, thereby limiting the contribution of additional NLP-based representations. In contrast, NN model assigns weight parameters across all input dimensions and can capture complex nonlinear interactions among features. This enables them to leverage the NLP features more effectively, leading to a pronounced enhancement in predictive performance. Furthermore, to assess model performance, we employ three error metrics: the coefficient of determination ($R^2$), mean absolute error (MAE), and mean squared error (MSE). Based on model performance on validation data, NN model is selected for further analysis. Note that while $R^2$ scores are shown in the parity plot, as they are more straightforward and interpretable (closer to 1 means better training), other metrics and parity plots for both models can be seen from supplementary information (Table S1).

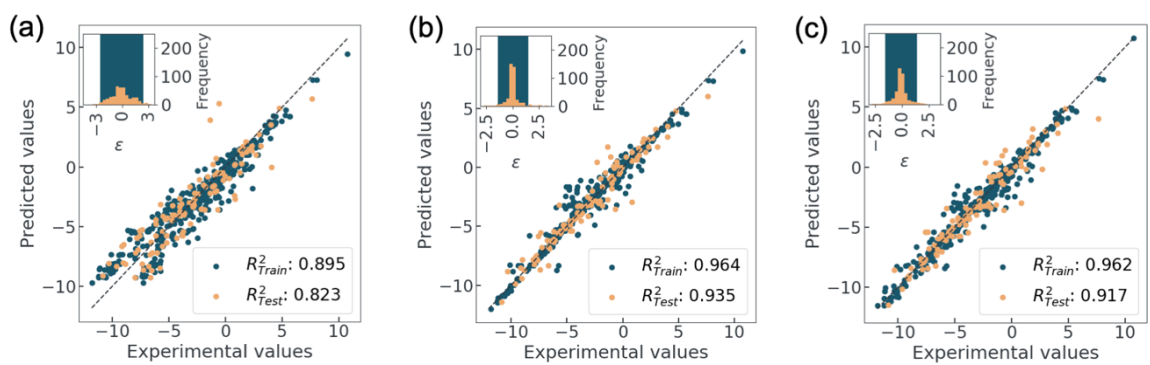

*Figure 4. Measured vs. predicted values of dissolution rate in log(dissolution)(g/m²/day) based on (a) purely composition-based model (b) composition-based NLP-ML model (c) the descriptor-based NLP-ML model using MLP for both the training and test datasets. The inset shows a histogram of errors, where the shaded region represents the 95% confidence interval.*

Figure 4(a) shows the parity plot of the MLP model performance without NLP features. We observe that the glass composition, pH and the temperature provide a reasonable prediction of the dissolution rate with $R^2$ score of 0.823. Figure 4(b) shows the performance of the NLP-ML model with composition-based inputs. We observe that the NLP-ML model (Fig. 4(b)) significantly outperforms the standard ML model confirming that the incorporation of the synthesis and testing conditions improve the predictive capability. It is noteworthy that the incorporation of NLP-derived embeddings provides more diverse and informative inputs, enriching the feature space and enables the NLP-ML model to exhibit superior performance in predicting the output variable, i.e., the dissolution rate. Furthermore, standard ML model performs poorly while predicting lower magnitude dissolution rates < $10^{-5}$ g/m²/day (see Fig. 1(a)). In contrast, the NLP-ML model performs well in extreme regions as well. Altogether, these results suggest that integrating natural language features enhances the model's predictive capability.

## 3.3. Descriptor-based Natural Language Processing-enhanced Machine Learning (NLP-ML) Model

While NLP-ML model provides better predictive capability, they are limited to glasses with components that are already included in the training data; a well-known limitation associated with composition-based inputs. To develop a more generalized predictive model that generalizes to unseen components that are absent in the training data, we perform feature engineering to develop periodic table-based descriptors that can be obtained by simple transformations of the glass compositions[27,76,77]. Thus, for the descriptor-based NLP-ML model, we consider features based on physical and chemical descriptors, pH, temperature and textual features as the input variables, and dissolution rate as the output variable of the model training (see Methods) for details.

Figure 4(c) shows that the performance of the descriptor-based NLP-ML model (Fig. 4(c)) is comparable with that of the composition-based NLP-ML model (Fig. 4(b)), demonstrating that the selected physical and chemical descriptors indeed efficiently capture the glass structure information. Furthermore, the descriptor-based NLP-ML model (Fig. 4(c)) outperforms the standard model (Fig. 4(a)) on both training and test sets which illustrates the efficacy of combining the text-based data modality with the numerical features to improve model inference. More details regarding the descriptors and the ML methodology used in this study, can be found in the Methods section.

### 3.4. Model Evaluation

To evaluate the descriptor-based NLP-ML model's generalizability, we curated a small dataset consisting of 32 data points, extracted from the published works of Fournier et al.[64] and Inagaki et al.[65] comprising silicon initial dissolution rate data with respect to International Simple Glass (ISG) (Fournier et al.[64]) and a Japanese simulated high-level waste glass P0798 (Inagaki et al.[65]). For more information regarding this curated dataset, refer the Methods Section. We employ the Jaccard distance metric as a quantitative measure to characterize the compositional differences between this dataset and the training data.

Figure 5(a) shows the Jaccard distances between glass compositions used in the training and test set with their closest neighbor in the training data. The vertical lines represent the distance values between the ISG and the P0798 waste glass and their closest neighbors in the training data. The Jaccard Distance value for ISG suggests that the training data consists of glasses with the same chemical components. However, in the case of the P0798 Glass, it can be inferred that the model has not encountered any glass with its chemical complexity during its training process. This is expected as the P0798 Glass consists of 34 chemical components, which exceeds the maximum chemical complexity (i.e., 28 chemical components) of all glasses present in the training and test data. It is also worth noting that even though, the feature space of the training and test set consists of 53 unique compositional features, the P0798 glass consists of chemical components which are not present in either dataset. To see the exact chemical composition of the glasses used in this dataset, refer supplementary information Tables S2-S3. Therefore, this curated dataset provides a fair test for the extrapolation capabilities of the descriptor-based NLP-ML model towards new glass compositions.

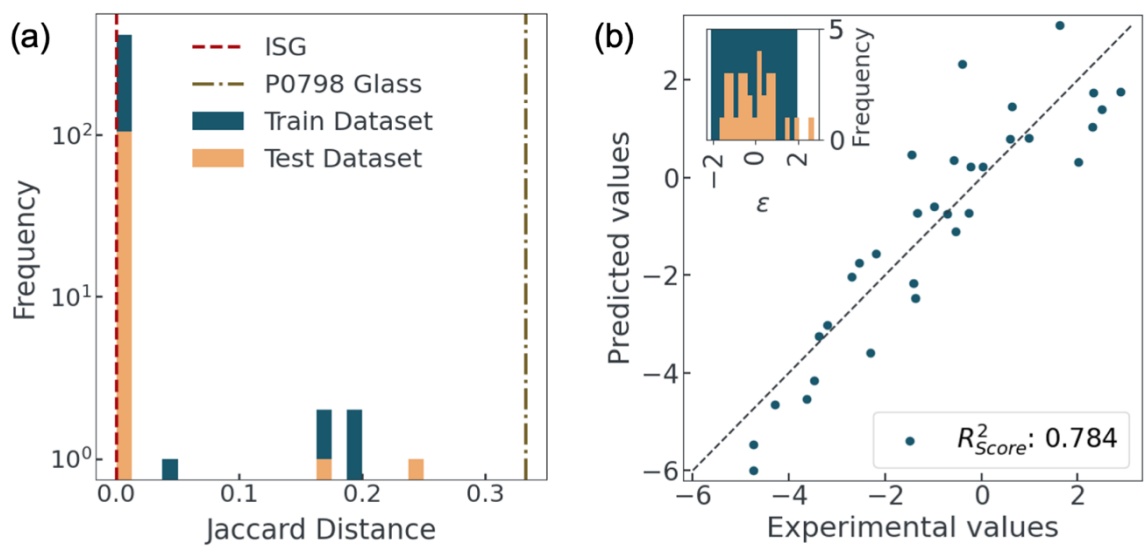

***Figure 5.*** *(a) Plot depicting the distribution of Jaccard Distance between the closest neighbours in the Training and Test sets with the vertical lines representing the Jaccard Distance values between ISG and P0798 Glass and their closest neighbours in the training data. (b) Performance of the descriptor-based NLP-ML model on the curated unseen data (Fournier et al.[64] and Inagaki et al.[65]). The inset shows a histogram of errors, where the shaded region represents the 95% confidence interval.*

Figure 5(b) shows the parity plot of the descriptor-based NLP-ML model performance on the unseen dataset, wherein the inset shows a histogram of errors with the shaded region representing the 95% confidence interval. The parity plot indicates that, owing to the generalizable descriptor-based methodology as well as training with textual data, the model has learned a more detailed characterization of the output variable, i.e. the dissolution rate. Notably, the model can predict the leaching rates of both ISG and the P0798 Glass with reasonable accuracy as highlighted by the $R^2$ scores (0.784; see Fig. 5(b)). These results suggest that the predictions are based on systematically learned patterns rather than loosely defined heuristic or random inference. These results also suggest that it can be inferred that the proposed ML methodology could serve as a reliable method for developing highly accurate prediction models for estimating the dissolution rate of glasses and modelling glass dissolution behavior. Moreover, this highlights the potential for developing NLP-augmented machine learning models to predict complex properties such as viscosity, glass-forming ability etc.

### 3.5. SHAP for descriptor-based NLP-ML model

We analyze the composition-based and descriptor-based NLP-ML models' predictions to understand the features that substantially impact the predictive performance of the models. Figure 6(a) shows the bee-swarm plot of the composition-based NLP-ML model, highlighting the top 20 features that most strongly influence model predictions, with the color map indicating the normalized feature value (see supplementary Information Fig. 25(a) for the corresponding bar plot). Notably, several features derived from textual embeddings rank among the most important, alongside composition, pH and temperature, underscoring the significant contribution of NLP-derived features in predicting dissolution rates. Experimental parameters such as temperature and pH emerge as highly influential, consistent with expectations and previous literature, reflecting their critical role in governing the thermodynamics of elemental leaching. This is because, in the dissolution dataset, we have data with a broad temperature (4°C to 300°C) and pH (1-13) range; hence, the model also learns the impact of temperature and pH variability. It can also be observed that compositional features including $SiO_2$ and $B_2O_3$, which are important network formers and $Na_2O$ and CaO, which are prominent network modifiers, significantly impact the dissolution rate predictions of the model. The Shapley values indicate that a higher quantity of $SiO_2$ drives the model to predict a lower dissolution rate value, which is consistent with the experimental observations of reduced leaching due to the formation of a silica-rich gel layer. Similarly, a higher quantity of $Na_2O$ and CaO drives the model to predict a higher dissolution rate. $Fe_2O_3$ and FeO seem to lack a straightforward relationship with the model predictions.

Figure 6(b) displays the bee-swarm plot for the descriptor-based NLP-ML model (see supplementary Information Fig. 25(b) for corresponding bar plot). It shows the top 20 features that played a crucial role in the descriptor-based NLP-ML model's performance, with the color map indicating the normalized feature value. Note that the most important feature of model performance is *temperature* as well. It can be observed that the effect of pH and temperature on the model output is positively correlated. Like the case of the composition-based NLP-ML model, higher magnitude of these features, result in a higher value of predicted dissolution rate and vice versa. It is worth noting that several important features that govern the predictions of the descriptor-based NLP-ML model are the natural language input features ($f_n$, n = 1, 2, 3, ....,10). This ascertains the benefits of adding testing processing conditions in addition to the composition, pH, and temperature. However, the interpretability of the descriptor-based NLP-ML model's features is limited, as individual natural language input features lack explicit meaning. Instead, they collectively represent the overall methodologies of experimentation and the testing conditions. On the contrary, if the model were to have been more interpretable, the use of the text-based modality would have had to be restricted, consequently affecting the performance of the NLP-ML model.

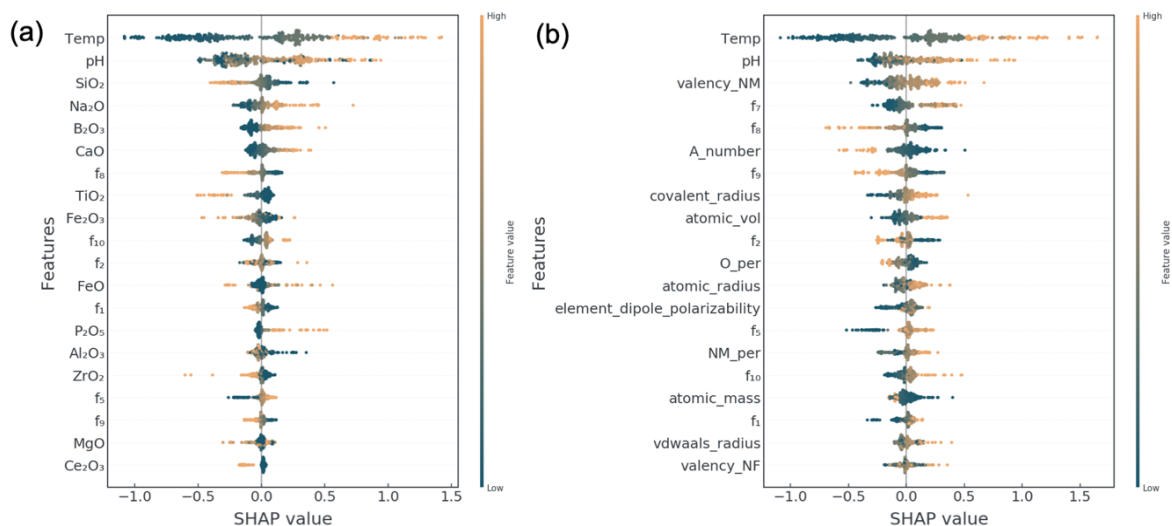

**Figure 6.** *SHAP bee-swarm plots for the (a) composition-based NLP-ML model (b) descriptor-based NLP-ML model displaying the important features in order of priority (most important to least important).*

## 4. CONCLUSIONS

Altogether, we proposed a framework to incorporate unstructured textual descriptions, associated with the synthesis and testing conditions, to predict the kinetics of glass dissolution. Specifically, with the NLP-ML model, we observed that the integration of textual data with numerical features results in substantial enhancements in predicted accuracy. Following this, we demonstrated that integration of textual data with periodic table-based descriptors result in a transferable and generalizable dissolution rate prediction model, enabling generalization to glass compositions with components that are not present in the training data. Finally, interpreting these models with SHAP offered valuable insights into the contributions of specific input features, clarifying the complex relationships between composition and dissolution rates. The framework presented in this study enables robust and generalizable predictions of glass dissolution while also providing a fillip toward predicting material properties that are notably affected by the processing and testing conditions.

**Declaration of Competing Interest**

The authors declare that they have no known competing financial interests or personal relationships that could have appeared to influence the work reported in this paper.

**Acknowledgments**

N.M.A.K. acknowledges the financial support for this research provided by the Department of Science and Technology, India, under the INSPIRE faculty scheme (DST/INSPIRE/04/2016/002774) BRNS YSRA (53/20/01/2021-BRNS), DST SERB Early Career Award (ECR/2018/002228) award and ANRF ARG by the Government of India. S. Mannan acknowledges the funding support from the Prime Minister's Research Fellowship (PMRF), Ministry of Education, Government of India. The authors thank the IIT Delhi HPC facility for providing computational and storage resources.

# Supplementary Information

| MODEL | | R² SCORE | | MAE | | MSE | |
|---|---|---|---|---|---|---|---|
| | | Train | Test | Train | Test | Train | Test |
| XGBoost | descriptor (with NLP) | 0.943 | 0.871 | 0.678 | 0.893 | 0.810 | 1.764 |
| | composition (with NLP) | 0.947 | 0.874 | 0.641 | 0.904 | 0.758 | 1.722 |
| | composition (No NLP) | 0.896 | 0.831 | 0.915 | 1.101 | 1.471 | 2.311 |
| ANN | descriptor (with NLP) | 0.962 | 0.917 | 0.486 | 0.737 | 0.535 | 1.135 |
| | composition (with NLP) | 0.964 | 0.935 | 0.406 | 0.629 | 0.515 | 0.88 |
| | composition (No NLP) | 0.895 | 0.823 | 0.955 | 1.174 | 1.496 | 2.420 |

*Table S1.* Comparison of XGBoost and ANN machine learning algorithms with the display of error metrics: $R^2$ score, Mean Absolute Error (MAE) and Mean Squared Error (MSE). Models shaded in colours are descriptor-based NLP-ML (blue) and composition-based NLP-ML (yellow) presented in the study.

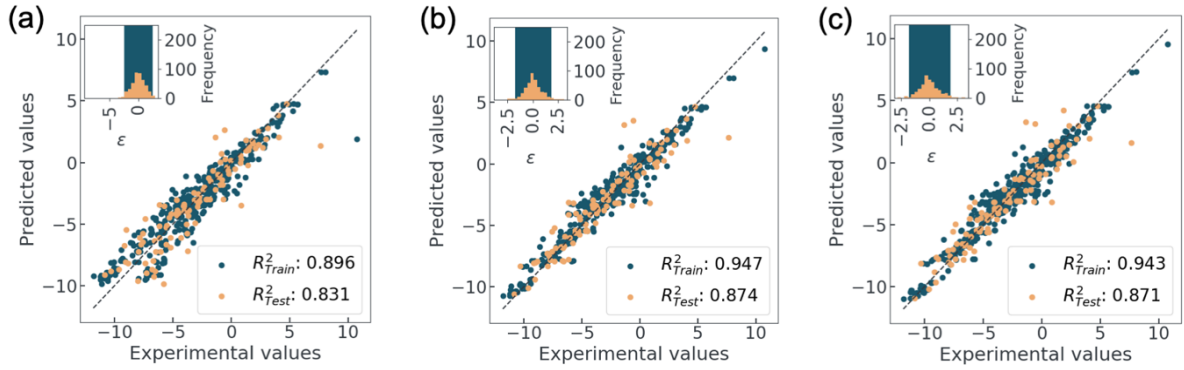

***Figure. 1.*** *Measured vs. Predicted values of dissolution rate in log(dissolution)(g/m$^2$/day) of (a) the standard vanilla model (b) the composition-based model with NLP features (c) the descriptor-based model with NLP features using XGBoost ML models for both the training and test datasets. The inset shows a histogram of errors, where the shaded region represents the 95% confidence interval.*

| | |
|---|---|
| $SiO_2$ | 56.2 |
| $B_2O_3$ | 17.3 |
| $Na_2O$ | 12.2 |
| $Al_2O_3$ | 6.1 |
| CaO | 5.0 |
| $ZrO_2$ | 3.3 |

***Table S2.*** *Composition of ISG from reference Fournier et al.[1]*

| | | | | | |
|---|---|---|---|---|---|
| $SiO_2$ | 46.60 | $Cs_2O$ | 0.75 | **$SeO_2$** | 0.02 |
| $B_2O_3$ | 14.20 | SrO | 0.30 | $TeO_2$ | 0.19 |
| $Al_2O_3$ | 5.00 | BaO | 0.49 | $Y_2O_3$ | 0.18 |
| $Li_2O$ | 3.00 | $ZrO_2$ | 1.46 | $La_2O_3$ | 0.42 |
| CaO | 3.00 | $MoO_3$ | 1.45 | $CeO_2$ | 3.34 |
| ZnO | 3.00 | $MnO_2$ | 0.37 | $Pr_6O_{11}$ | 0.42 |
| $Na_2O$ | 10.00 | $RuO_2$ | 0.74 | $Nd_2O_3$ | 1.38 |
| $P_2O_5$ | 0.30 | **$Rh_2O_3$** | 0.14 | $Sm_2O_3$ | 0.29 |
| $Fe_2O_3$ | 2.04 | PdO | 0.35 | **$Eu_2O_3$** | 0.05 |
| NiO | 0.23 | $Ag_2O$ | 0.02 | $Gd_2O_3$ | 0.02 |
| $Cr_2O_3$ | 0.10 | CdO | 0.02 | | |
| $Rb_2O$ | 0.11 | $SnO_2$ | 0.02 | | |

*Table S3. Composition of P0798 Glass from reference Inagaki et al.[2]. Chemical components highlighted in bold are not present in the training and test datasets.*

| Name of hyperparameter | Range/List/Value |
|---|---|
| Epochs | 500 |
| n_layers | 2 to 6 |
| drop | True, False |
| drate | 0.1 to 0.4 (step=0.1) |
| norm | True, False |
| activation | LeakyReLU, ReLU |
| opt | Adam, SGD |
| lr(opt==SGD) | 1e-5 to 0.1 in log scale |
| momentum | 9e-5 to 0.9 in log scale |
| lr(opt==Adam) | 1e-3 to 0.1 in log scale |
| Weight_decay | 1e-5 to 1e-3 in log scale |
| Batch_size | [8, 16] |

*Table S4. Artificial Neural Network model hyperparameters.*

**Natural Language Input Features using Principal Component Analysis**

We also used Principal Component Analysis[3] (PCA), which is another dimensionality reduction technique, to reduce the large number of dimensions of the textual embeddings, to a select few principal components (or features). In this approach, dimensionality reduction is achieved using Singular Value Decomposition (SVD) of the data to project it onto a lower dimensional space. This is done for all 21 embeddings, each having a dimension of [n, 768]. Let $X$ be our vector embedding matrix:

$$X = \begin{bmatrix} a_{1,1} & \cdots & a_{1,768} \\ \vdots & \ddots & \vdots \\ a_{n,1} & \cdots & a_{n,768} \end{bmatrix}$$

Before performing Singular Value Decomposition on this matrix, the values are centered by subtracting the values with their column-wise mean.

*Column-wise mean for column 1 :*

$$\overline{a_1} = \frac{a_{1,1} + a_{2,1} + \cdots + a_{n,1}}{n}$$

Therefore, the resulting matrix becomes:

$$X_{centered} = \begin{bmatrix} a_{1,1} - \overline{a_1} & \cdots & a_{1,768} - \overline{a_{768}} \\ \vdots & \ddots & \vdots \\ a_{n,1} - \overline{a_1} & \cdots & a_{n,768} - \overline{a_{768}} \end{bmatrix}$$

Singular Value Decomposition:

$$X_{centered} = U\Sigma V^T$$

Here, the matrix $U$ is a $n \times n$ orthogonal matrix. The columns of this matrix form an orthonormal basis for the column space of $X_{centered}$. $V^T$ is a 768×768 orthogonal matrix and its rows form an orthonormal basis for the row space of $X_{centered}$. The rows of $V^T$ are the principal components. $\Sigma$ is a $n \times 768$ diagonal matrix and contains the singular values which quantify the amount of variance explained by each principal component. The columns of U, scaled by the singular values effectively represent the transformed datapoints in the new, lower dimensional space. The maximum number of non-zero singular values in $\Sigma$ is equal to the maximum value of the rank of $X_{centered}$ which cannot exceed **min(n, 768)**. Therefore, the 768 dimensions of the embeddings can only be transformed to a maximum of **min(n, 768)** principal components. Analysis was conducted regarding the appropriate number of textual features to be added by plotting the percentage of explained variance with increasing features for all 21 $n \times 768$ embeddings. Figures 2-22 display the plots of the percentage of explained variance with increasing features. The plots highlight the number of features that explain 95% of the variability of the textual data.

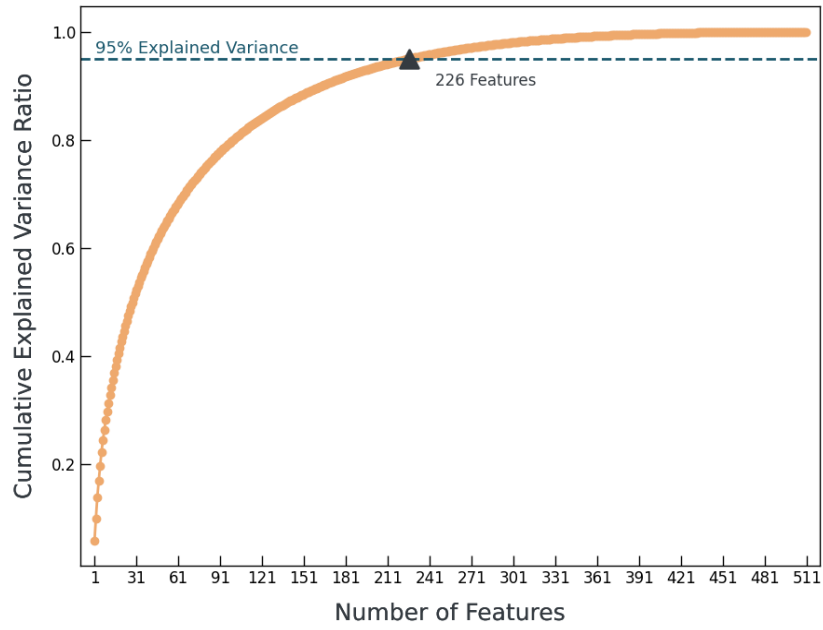

***Figure. 2.*** *Plot displaying the percentage of explained variance with increasing features for Embedding 1 (dimension - [510, 768])*

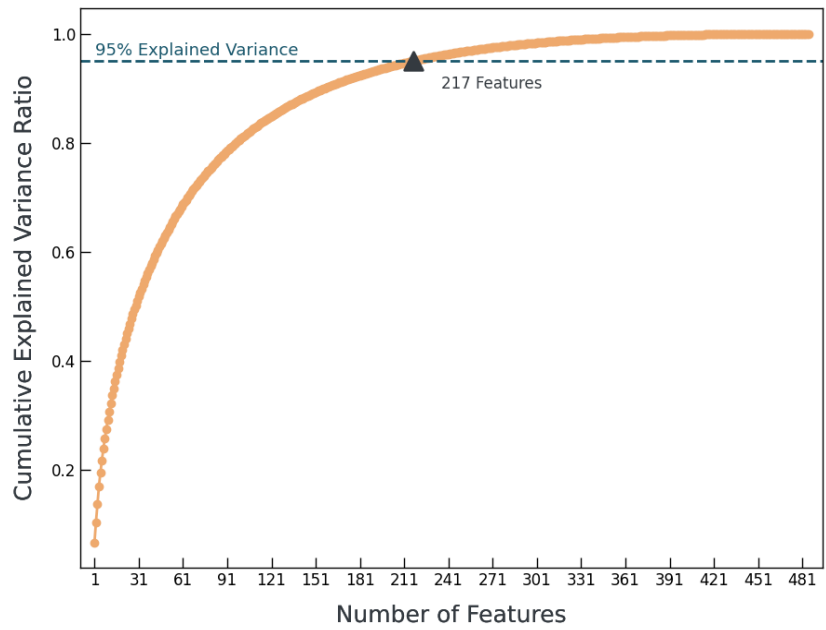

***Figure. 3.*** *Plot displaying the percentage of explained variance with increasing features for Embedding 2 (dimension - [485, 768])*

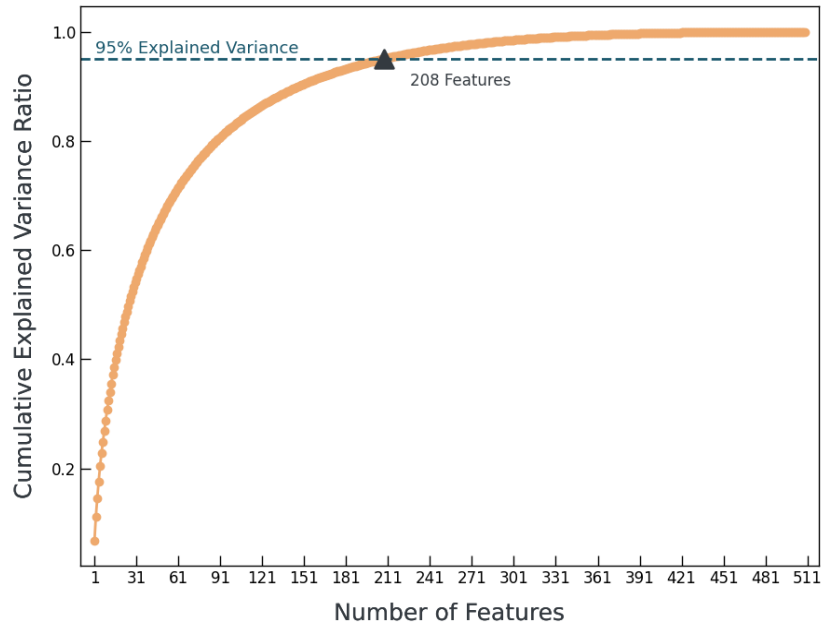

***Figure. 4.*** *Plot displaying the percentage of explained variance with increasing features for Embedding 3 (dimension - [509, 768])*

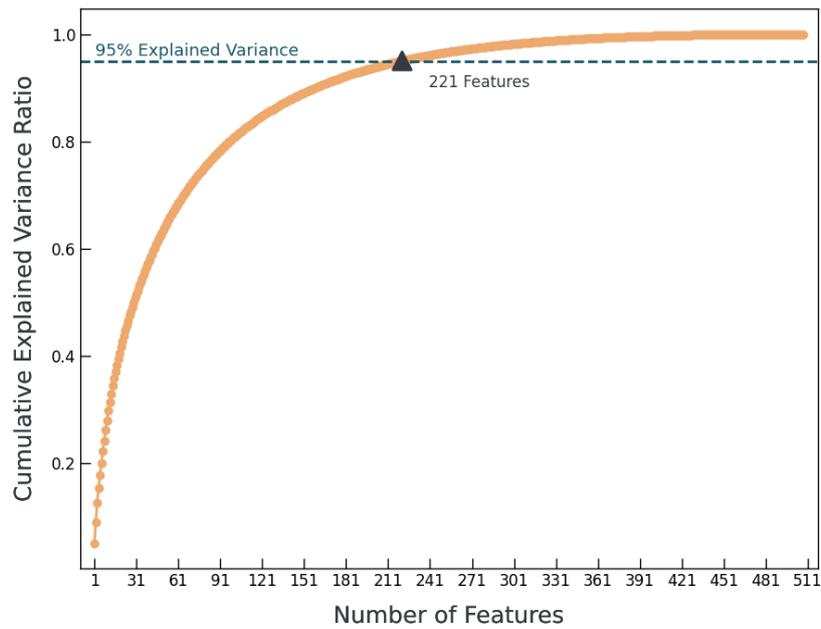

***Figure. 5.*** *Plot displaying the percentage of explained variance with increasing features for Embedding 4 (dimension - [508, 768])*

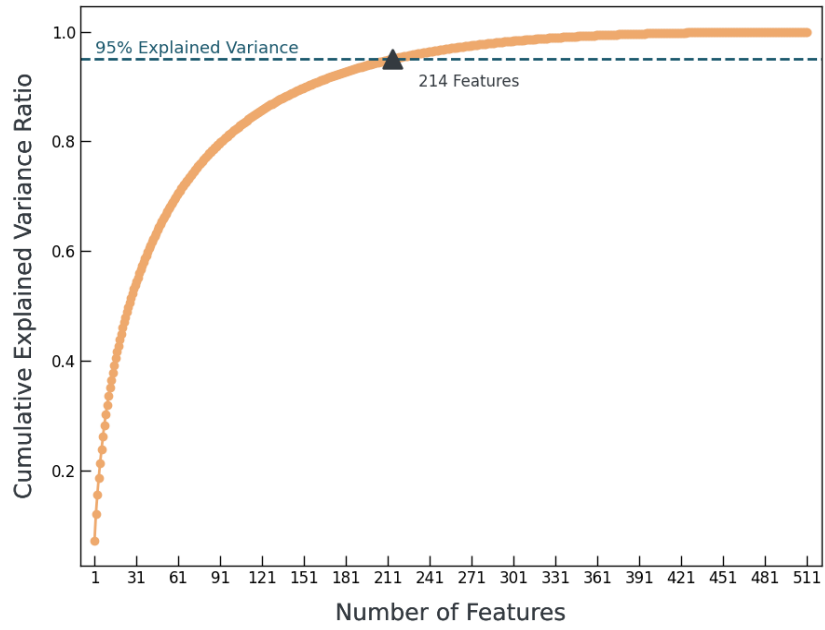

***Figure. 6.*** *Plot displaying the percentage of explained variance with increasing features for Embedding 5 (dimension - [511, 768])*

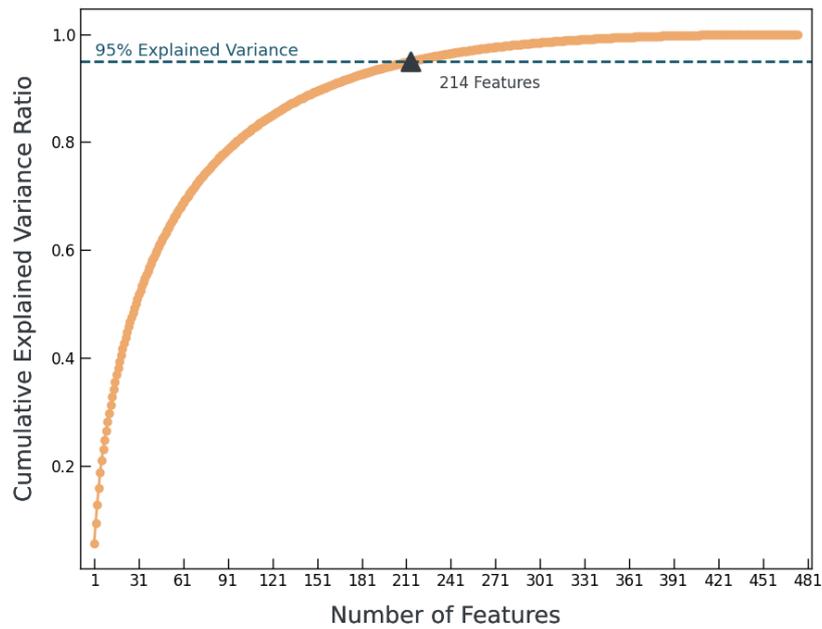

***Figure. 7.*** *Plot displaying the percentage of explained variance with increasing features for Embedding 6 (dimension - [474, 768])*

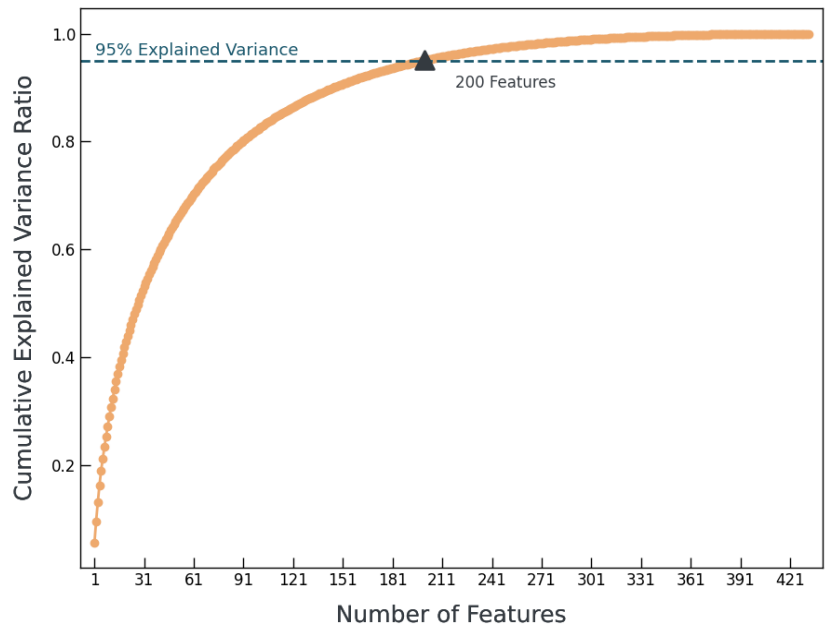

***Figure. 8.*** *Plot displaying the percentage of explained variance with increasing features for Embedding 7 (dimension - [432, 768])*

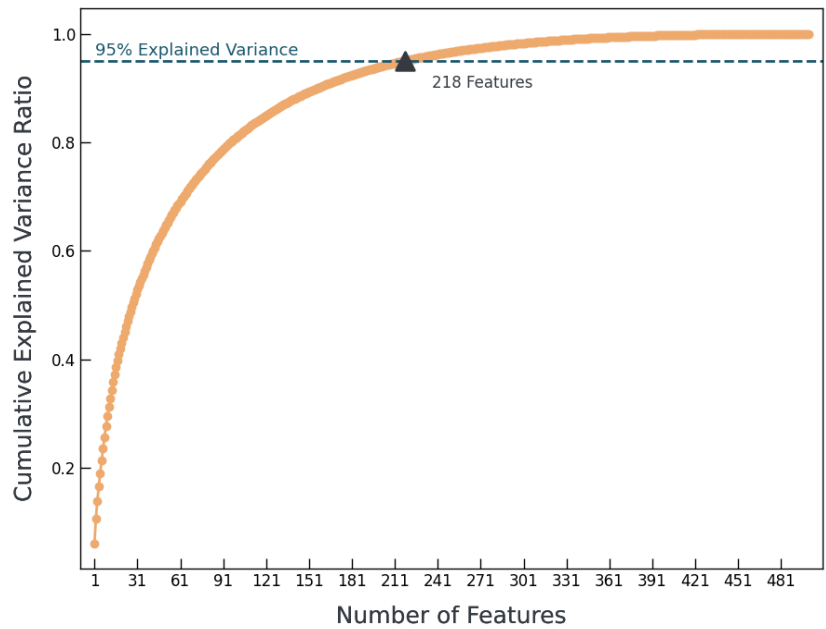

***Figure. 9.*** *Plot displaying the percentage of explained variance with increasing features for Embedding 8 (dimension - [500, 768])*

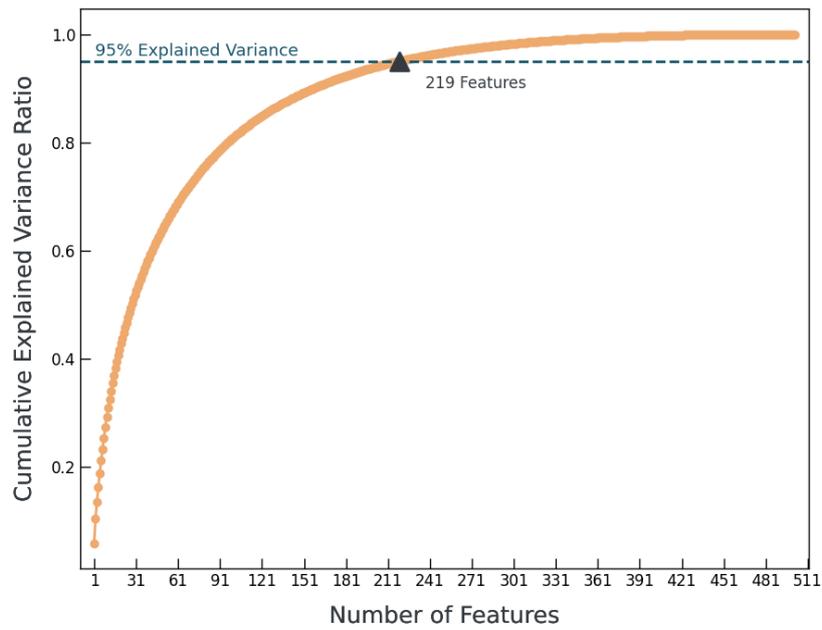

***Figure. 10.*** *Plot displaying the percentage of explained variance with increasing features for Embedding 9 (dimension - [502, 768])*

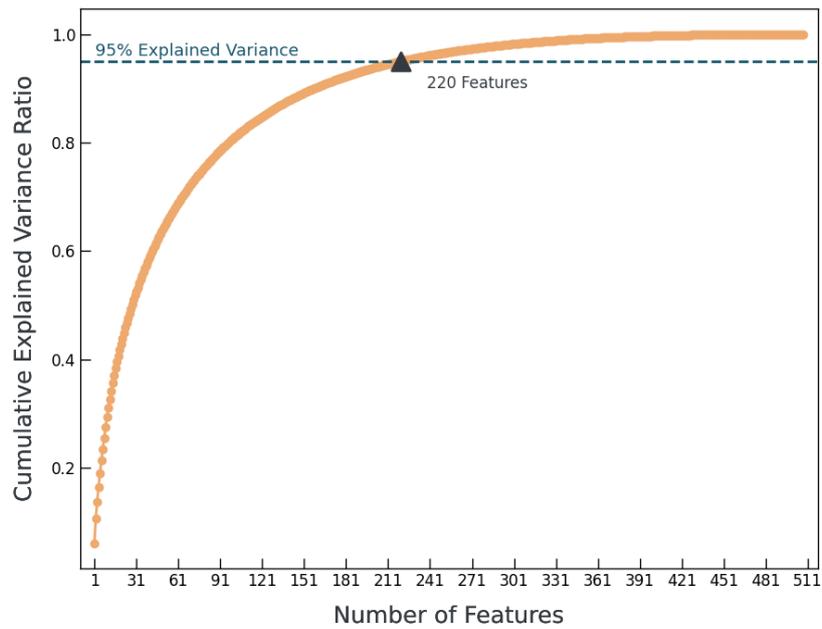

***Figure. 11.*** *Plot displaying the percentage of explained variance with increasing features for Embedding 10 (dimension - [508, 768])*

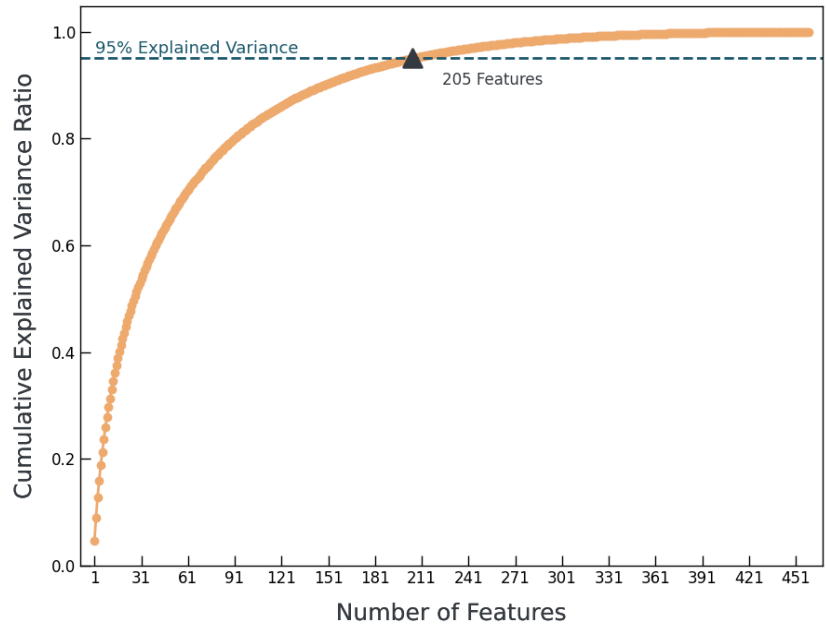

***Figure. 12.*** *Plot displaying the percentage of explained variance with increasing features for Embedding 11 (dimension - [459, 768])*

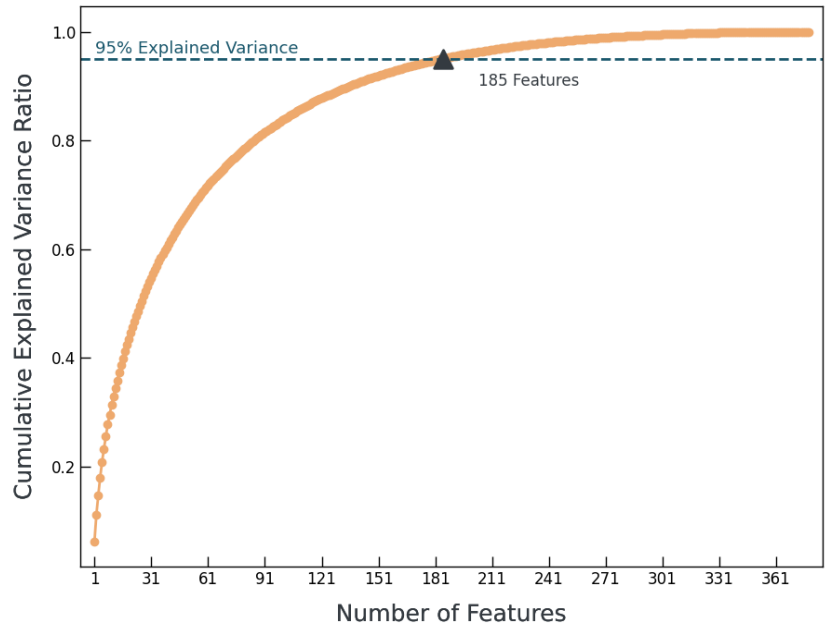

***Figure. 13.*** *Plot displaying the percentage of explained variance with increasing features for Embedding 12 (dimension - [378, 768])*

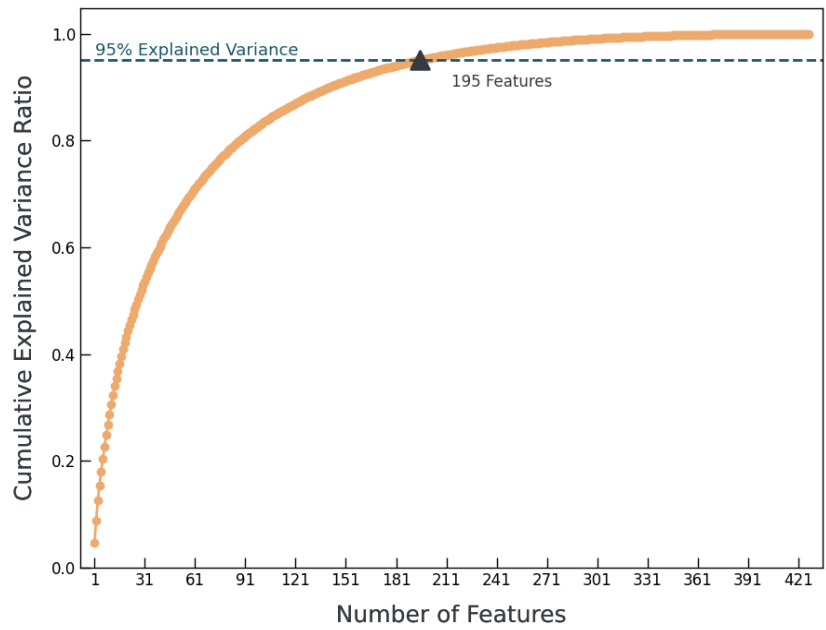

***Figure. 14.*** *Plot displaying the percentage of explained variance with increasing features for Embedding 13 (dimension - [427, 768])*

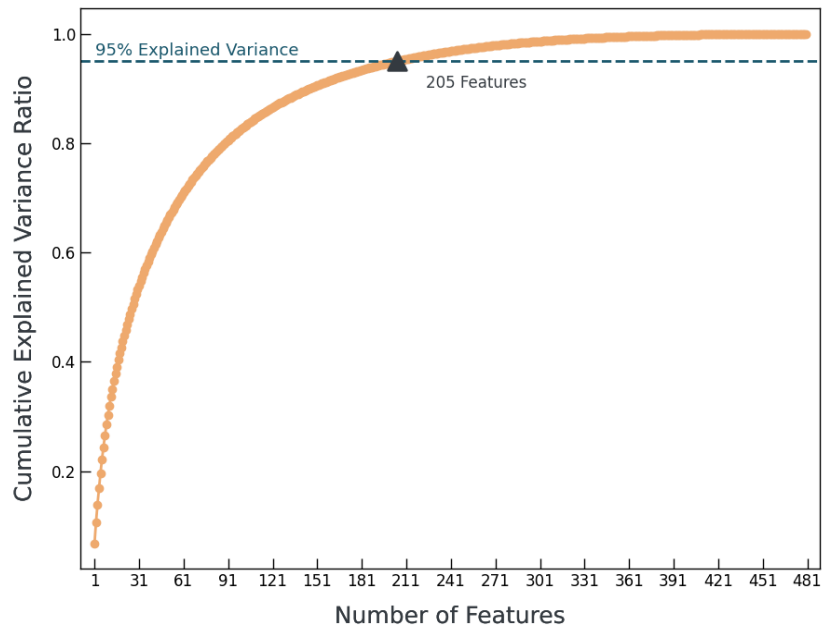

***Figure. 15.*** *Plot displaying the percentage of explained variance with increasing features for Embedding 14 (dimension - [480, 768])*

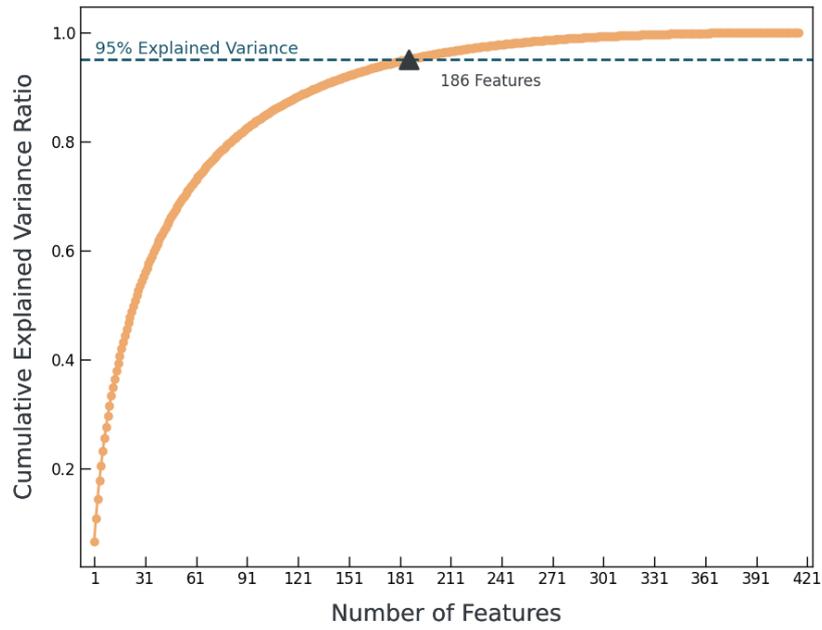

***Figure. 16.*** *Plot displaying the percentage of explained variance with increasing features for Embedding 15 (dimension - [416, 768])*

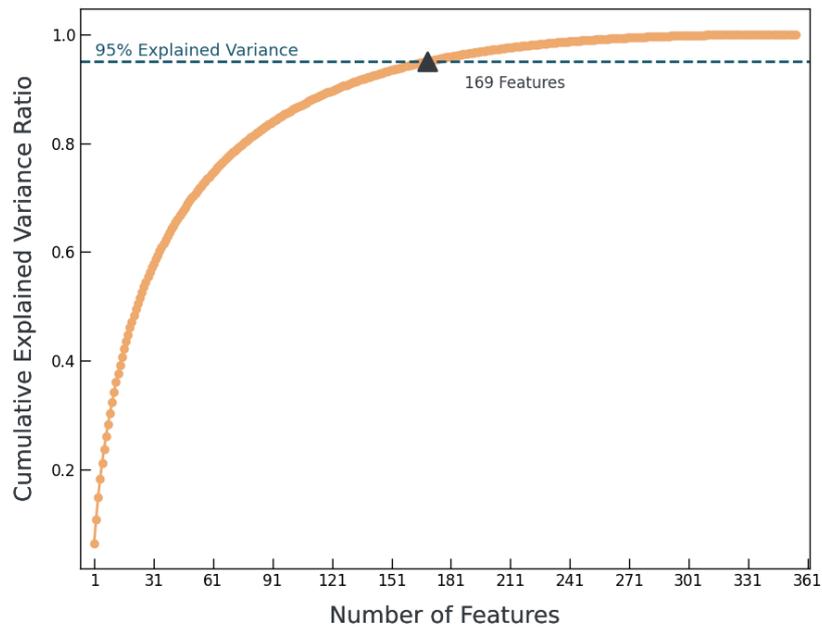

***Figure. 17.*** *Plot displaying the percentage of explained variance with increasing features for Embedding 16 (dimension - [355, 768])*

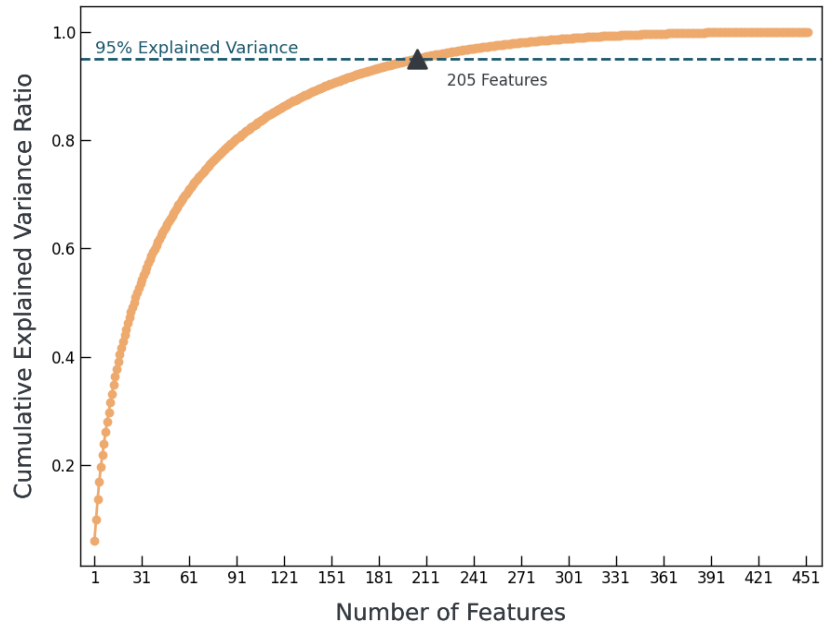

***Figure. 18.*** *Plot displaying the percentage of explained variance with increasing features for Embedding 17 (dimension - [452, 768])*

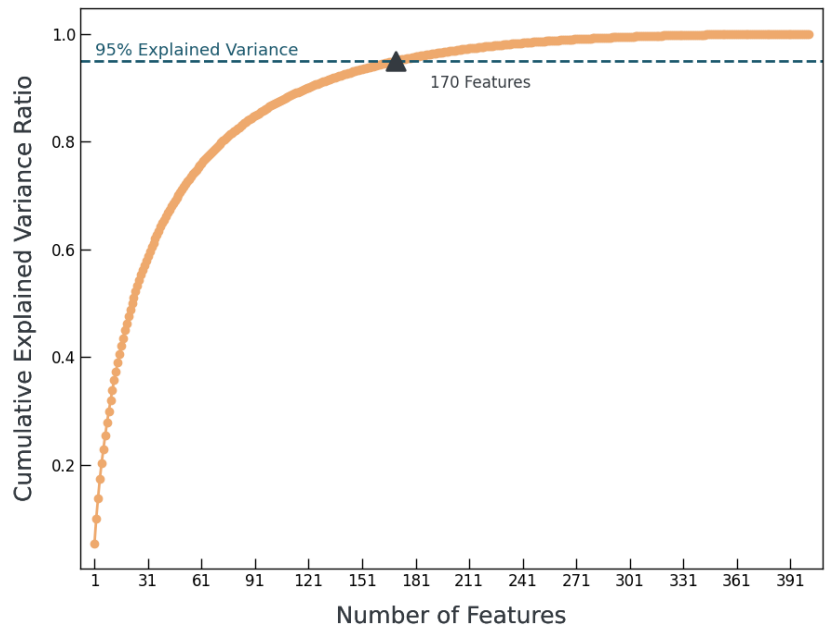

***Figure. 19.*** *Plot displaying the percentage of explained variance with increasing features for Embedding 18 (dimension - [401, 768])*

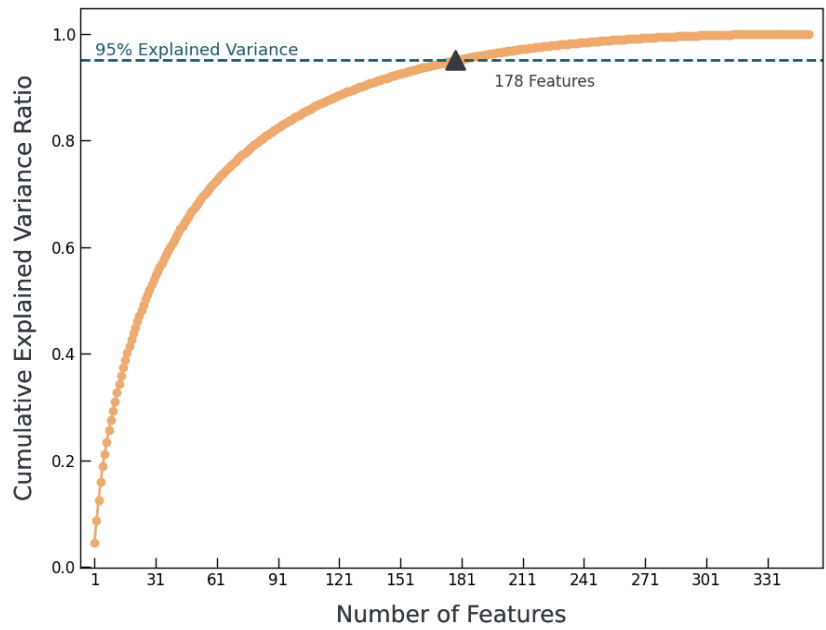

***Figure. 20.*** *Plot displaying the percentage of explained variance with increasing features for Embedding 19 (dimension - [351, 768])*

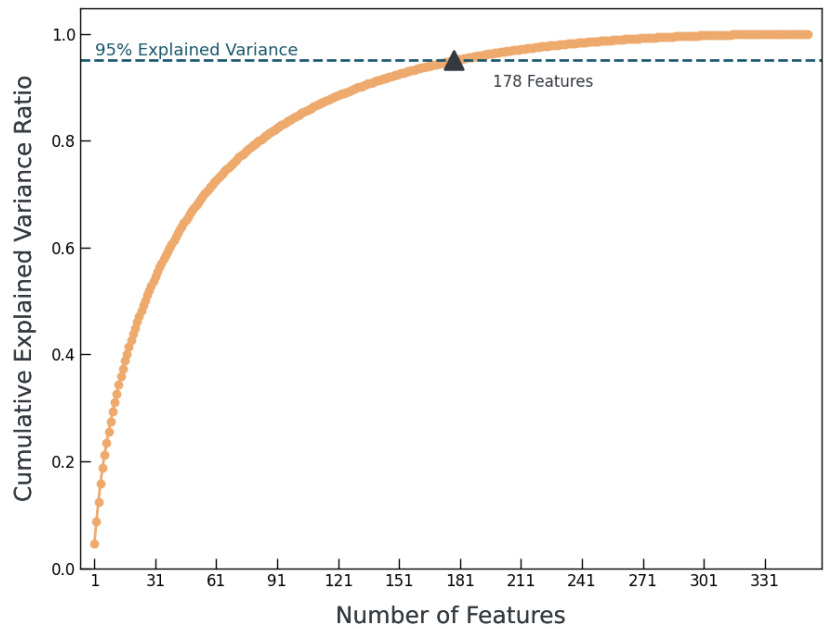

***Figure. 21.*** *Plot displaying the percentage of explained variance with increasing features for Embedding 20 (dimension - [352, 768])*

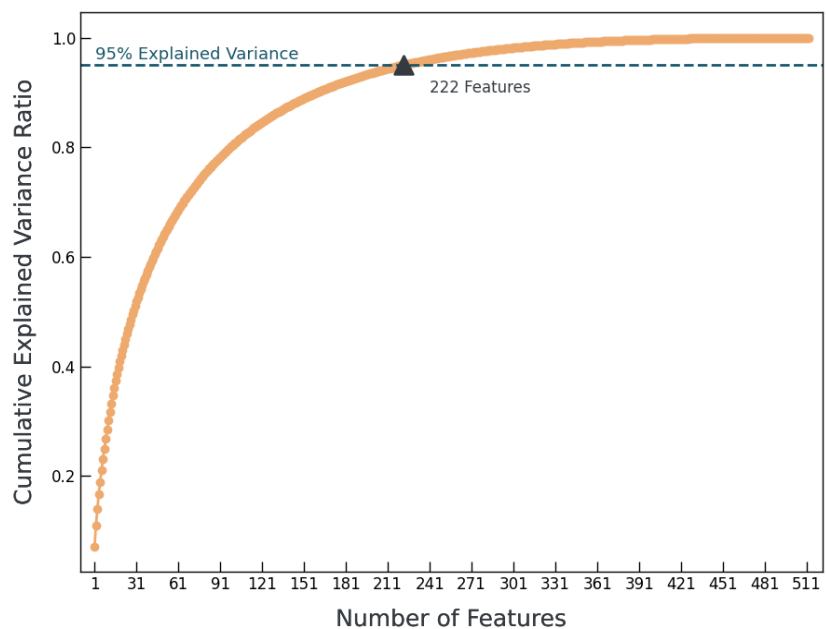

***Figure. 22.*** *Plot displaying the percentage of explained variance with increasing features for Embedding 21 (dimension - [512, 768])*

It can be observed from figures 2-22 that the minimum number of features that explain 95 % of the variance of the textual data for an embedding is 169 features. If the dataset were to incorporate the 169 additional features alongside the 53 compositional features, pH and temperature, it could lead to certain model training limitations. The compositional features, pH and temperature data could be overshadowed by the additional NLP features, potentially reducing predictive accuracy of the model. The model could develop a bias towards the NLP features and give less importance to the composition, pH and temperature, which could be counterproductive as glass properties like dissolution rate are a function of the glass composition and physical and chemical conditions of the surrounding medium. Therefore, similar to the reduction of the UMAP[4,5] features, the embeddings were reduced to 10 features/principal components using PCA. Fig. 23(a) displays the parity plot of the composition-based model trained with PCA-reduced textual features. Fig. 23(b) displays the parity plot of the descriptor-based model trained with PCA-reduced textual features. It can be seen that while these models only slightly underperform the composition-based NLP-ML and descriptor-based NLP-ML models respectively, Fig. 23(c), which displays the performance of the descriptor-based model trained with PCA-reduced textual features on the unseen data, highlights the lack in generalization capabilities of the model. Fig. 24(a) and Fig. 24(b) display the corresponding XGBoost models trained on PCA-reduced textual features. Fig. 24(a) displays the parity plot of the composition-based model and Fig. 24(b) displays the descriptor-based model. The inset of all plots shows a histogram of errors, where the shaded region represents the 95% confidence interval.

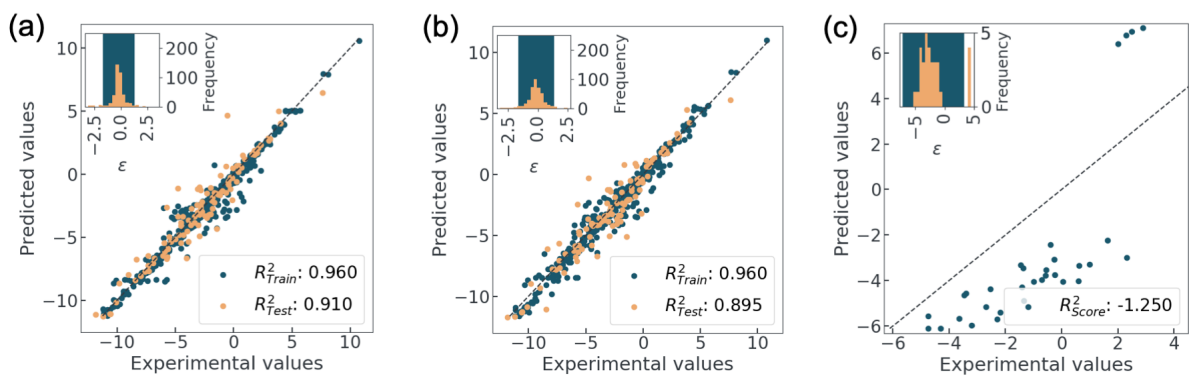

*Figure. 23. Measured vs. Predicted values of dissolution rate in log(dissolution)(g/m$^2$/day) of (a) composition-based model trained with PCA-reduced textual features (b) descriptor-based model trained with PCA-reduced textual features using ANN ML models for both the training and test datasets (c) Performance of the descriptor-based ANN model with PCA-reduced NLP features on the unseen data. The inset shows a histogram of errors, where the shaded region represents the 95% confidence interval.*

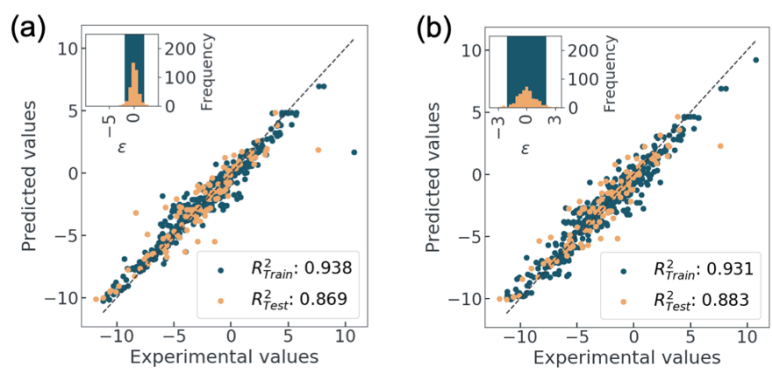

*Figure. 24. Measured vs. Predicted values of dissolution rate in log(dissolution)(g/m$^2$/day) of (a) composition-based model trained with PCA-reduced textual features (b) descriptor-based model trained with PCA-reduced textual features using XGBoost ML models for both the training and test datasets. The inset shows a histogram of errors, where the shaded region represents the 95% confidence interval.*

Fig. 25 displays the SHAP[6,7] bar plots for the model presented in this study. Note, bar plots give only the impact of features and doesn't tell anything about the directionality of impact.

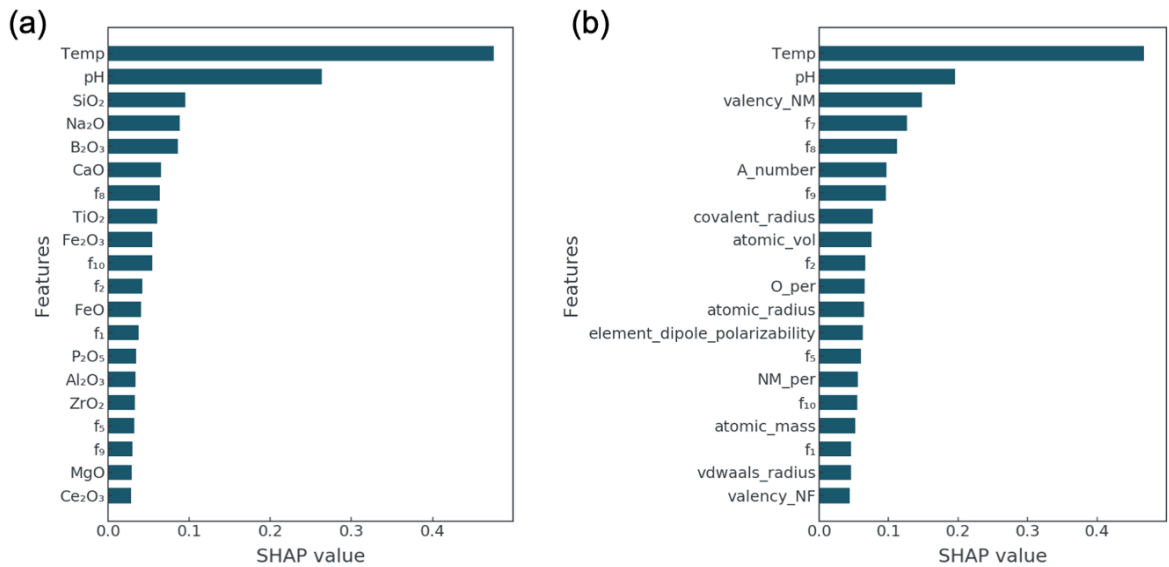

*Figure. 25. SHAP bar plots for the (a) composition-based NLP-ML model (b) descriptor-based NLP-ML model displaying the important features in order of priority (most important to least important).*